\documentclass[prl,nobalancelastpage,twocolumn,superscriptaddress,nolongbibliography]{revtex4-2}
\pdfoutput=1

\usepackage{amsmath,amssymb,amsfonts,mathtools}
\usepackage{graphicx}
\usepackage{xcolor}
\usepackage[colorlinks=true,citecolor=blue,linkcolor=blue,urlcolor=blue]{hyperref}
\usepackage{microtype}
\usepackage{overpic}

\newcommand{\prlsection}[1]{{\em {#1}.---~}}

\begin{document}

\title{Battery Locality Is Necessary in Noncommuting Quantum Charging Bounds}

\author{Sheryl Mathew}
\email{smathew@imsc.res.in}
\affiliation{Optics and Quantum Information Group, The Institute of Mathematical Sciences, CIT Campus, Taramani, Chennai 600113, India.}
\affiliation{Homi Bhabha National Institute, Training School Complex, Anushakti Nagar, Mumbai 400094, India.}

\begin{abstract}
Bounds on the charging power of quantum batteries in direct charging protocols are often interpreted as charging-side constraints. Although the locality of the charging Hamiltonian is known to restrict attainable power in certain protocols, whether battery locality is itself operationally necessary has remained unresolved. Here we show that it is necessary for spin batteries whose interactions do not pairwise commute. We construct commuting and noncommuting high-locality batteries with identical interaction supports and local energy scale, driven by the same charging. The commuting battery saturates the battery-locality-independent bound, whereas the noncommuting battery exceeds it by a parametrically growing factor, demonstrating that this bound cannot apply generally. The enhanced norm of the commutator is dynamically attained from a suitable joint state supported entirely in the battery ground-energy sector, although this state may be correlated with the auxiliary fermionic degrees of freedom. Our results establish battery locality as an operational resource enabled by noncommuting battery interactions.
\end{abstract}

\maketitle

\prlsection{Introduction} Quantum batteries can offer a quadratic advantage in charging power over conventional parallel charging, whose power scales only linearly with the number of sites or local cells \cite{Binder2015,Campaioli2017}. Early efforts to understand this advantage emphasized the possible role of the interaction order of the charging Hamiltonian and investigated how attainable power may be bounded by charging-side constraints \cite{Campaioli2017,Gyhm2022,Campaioli2024,Rossini2020}. Much of this framework assumed non-interacting or effectively locally decomposable batteries \cite{Binder2015,Campaioli2017,Gyhm2022}.

Interactions within the battery are now known to influence charging and storage through their strength and range \cite{Le2018}, frustration and boundary conditions \cite{Catalano2024,Zheng2025}, spectral structure near quantum phase transitions \cite{Grazi2024,Farina2026}, and microscopic features such as counterrotating terms \cite{Chen2026}. The topology and chirality of the battery coupling graph \cite{Cavazzoni2026}, as well as atomic interactions and long-range spin couplings in charging-mediated architectures \cite{Dou2022Dicke,Dou2022Heisenberg}, can likewise modify charging power and storage capacity. These results establish that the structure of the battery Hamiltonian matters dynamically. They do not, however, determine whether the interaction locality of the battery is itself necessary in a general bound on charging power.

In a direct-charging protocol, the quantum battery is initially prepared in a state of low energy with respect to its Hamiltonian $H_B$ and subjected, at time $t=0$, to a sudden quench to a charging Hamiltonian $H_C$ \cite{Campaioli2017,Le2018,Rossini2020}. The state evolves under $H_C$ until a time $t_C$, at which point the system is quenched back to $H_B$. The aim is to reach a state of higher energy with respect to $H_B$. The instantaneous charging power is
\begin{equation}
P(t)=\frac{d}{dt}\langle H_B\rangle_t
= i\langle[H_C,H_B]\rangle_t,
\label{eq:power}
\end{equation}
so the central object is the joint algebra of the battery Hamiltonian $H_B$ and the charging Hamiltonian $H_C$ \cite{SarkarGhosh2025}.

Most relevant to the present work are recent rigorous results showing that, for general non-commuting Hamiltonians, the commutator can depend on the localities of both Hamiltonians \cite{SarkarGhosh2025,Kuwahara2016}. In contrast, for a commuting battery, the rigorous bound on charging power does not depend on battery locality \cite{Gyhm2022,Kuwahara2016,AradKuwaharaLandau2016}. The appearance of battery locality in the general non-commuting bound does not, by itself, establish that this dependence is operationally necessary: it could, in principle, be an artifact of the bounding method, with a stronger locality-independent bound remaining valid.

Here we show that this is not the case. We establish the decisive role of the battery locality $k_B$ using an exactly solvable battery defined on a complete graph. We construct commuting and non-commuting batteries with the same interaction-support geometry, locality, and local energy scale, and subject them to the same charging Hamiltonian. The commuting construction saturates the corresponding $k_B-$independent bound, whereas the non-commuting construction exceeds it by a parametrically growing factor. Our results therefore show that the dependence on $k_B$ is not merely an artifact of the general upper bound, but can govern the scaling of the maximum attainable charging power. Since the commuting-case upper bound is exceeded while the relevant local-energy and charging-resource constraints are held fixed, no commuting battery can achieve the same enhancement under those constraints. We further demonstrate that a suitable joint state supported entirely in the battery ground-energy sector dynamically attains the enhanced commutator norm. The required joint state may be correlated with auxiliary fermionic degrees of freedom; thus, the result establishes dynamical attainability when such correlated initialization is allowed.

\prlsection{Commuting versus non-commuting locality bounds}

Consider a lattice $\Lambda$ of $N$ sites, $i=1,\ldots,N$, with local Hilbert-space dimensions $d_i$, which we take here to be $d_i=2$ for our spin models. We write a Hamiltonian $H$ as a sum of interaction terms,
\begin{equation}
    H = \sum_{X\subseteq\Lambda} h_X,
\end{equation}
where $h_X$ is supported on the subset $X\subseteq\Lambda$.

\begin{figure*}[t!]
    \centering
    \includegraphics[width=\textwidth]{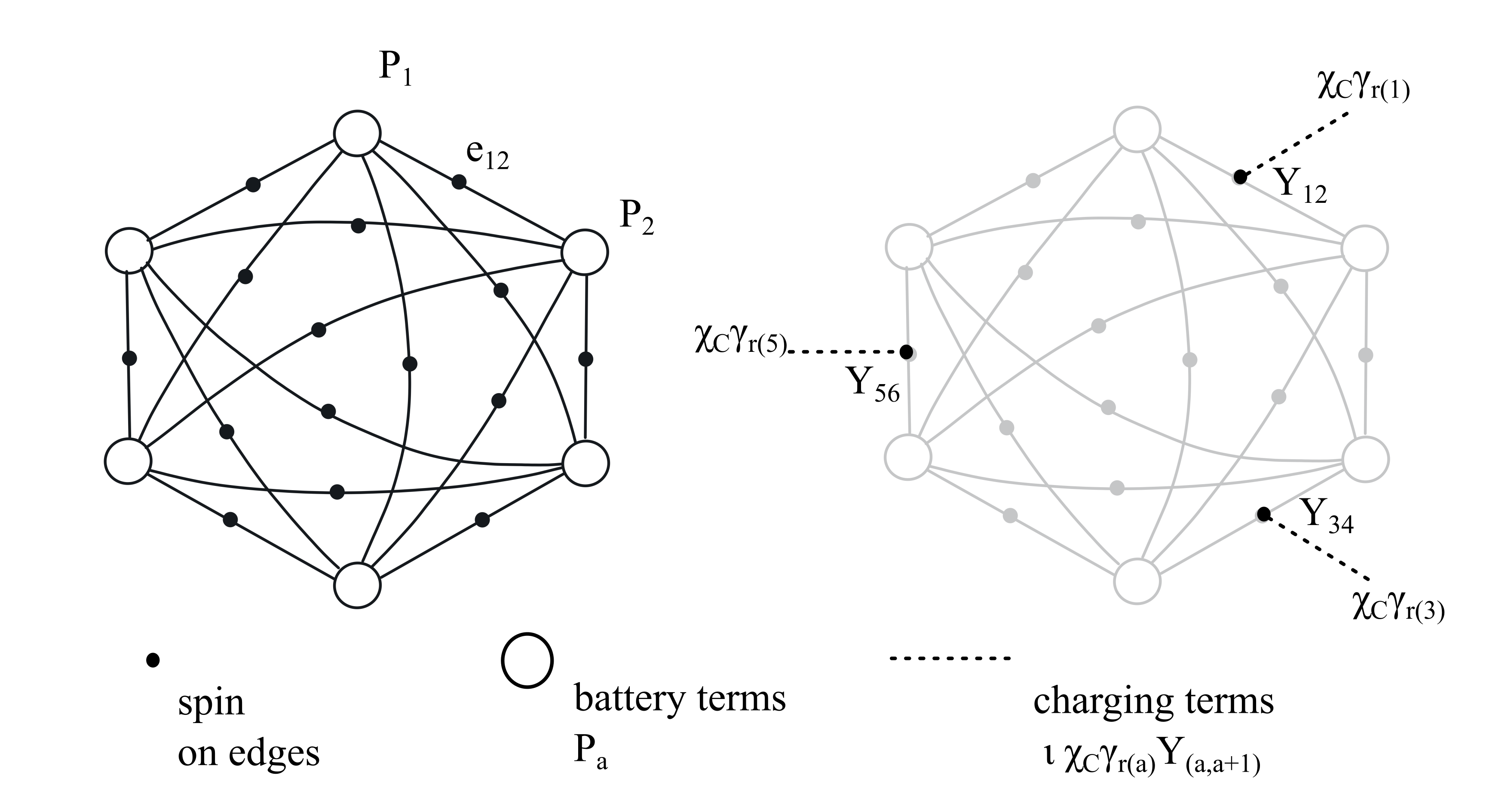}
    \caption{Battery (left) and charging (right) Hamiltonian geometry for $M=6$. Spins reside on the edges $e_{ab}$ of the complete graph $K_M$. Each vertex $a$ labels a battery interaction string $P_a$, supported on the $M-1$ incident edge spins. For the charging, a perfect matching $\mathcal{M}=\{(1,2),(3,4),(5,6)\}$ is selected. Each matched edge $e_r$ is acted on by $Y_{e_r}$ and coupled to an anticommuting central fermionic channel $\chi_C\gamma_{r}$ (shown schematically by dashed lines). Unselected battery edges are faded in the charging schematic.
    }
    \label{fig:K6_hamiltonian_geometry}
\end{figure*}

The $k$-locality, or interaction order, of $H$ is the maximum support size of any nonzero interaction term,
\begin{equation}
    k = \max_{X:h_X\neq 0} |X|.
\end{equation}
Further, the $g$-extensivity is defined as the maximum cumulative interaction strength incident on any site,
\begin{equation}
    g = \max_{i\in\Lambda}
    \sum_{X\ni i} \|h_X\|.
\end{equation}

The AKLH lemma \cite{AradKuwaharaLandau2016}, and its adaptation to the quantum-battery context \cite{SarkarGhosh2025}, establishes a strong restriction for a commuting battery Hamiltonian $H_B$ (all of whose interactions $h^B_X$ and $h^{B}_{X^{'}}$ commute pairwise, $\|[h^B_X,h^{B}_{X^{'}}]\| = 0$) on the energy separations that can be coupled by the charging Hamiltonian $H_C$. In particular,
\begin{equation}
    \Pi_{E'} H_C \Pi_E = 0
    \qquad
    \text{for}
    \qquad
    |E'-E| > 2 g_B k_C,
\end{equation}
where $\Pi_E$ denotes the projector onto the eigenspace of $H_B$ with energy $E$.

Thus, the charging-power bound of \cite{Gyhm2022}, which depends on the separation $\Delta E_B$ such that $\Pi_{E+\Delta E_B} H_C \Pi_E \neq 0$,
\begin{equation}
    |P(t)| \leq \Delta E_B \|H_C\|,
\end{equation}
together with an accessible energy width
\begin{equation}
    \Delta E_B \simeq 2 k_C g_B,
\end{equation}
leads to an upper bound on the maximum instantaneous power:
\begin{equation}
    |P(t)| \leq 2 k_C g_B \|H_C\|.
\end{equation}
This bound (up to constant prefactors) was proven in \cite{Kuwahara2016, SarkarGhosh2025}. In the quantum battery context, similar bounds accounting solely for the charging Hamiltonian's locality have been proven in \cite{Gyhm2022, Campaioli2017}.

For the more general non-commuting case, \cite{Kuwahara2016} established the rigorous bound which depended on the localities of both operators appearing within the commutator, and which was subsequently introduced into the context of quantum batteries (upto constant prefactors) in \cite{SarkarGhosh2025}:
\begin{equation}
    |P(t)|
    \leq
    6 k_B k_C g_B \|H_C\|,
\end{equation}
For completeness, we refer to the Supplemental Material \cite{SupplementalMaterial} for the relevant derivations.

Specifically, we note that if only the part of $H_C$ that is incident on the support of the battery Hamiltonian $H_B$ contributes to the commutator, its relevant interaction order may be replaced by an effective locality
\begin{equation}
    k_C^{\mathrm{eff}} \leq k_C.
\end{equation}

\prlsection{Explicit realization of non-commuting bound}
We define the interactions and lattice sites of $H_B$ on the vertices and edges, respectively, of a complete graph $K_M$, with $M$ even. Given vertex labels $a,b$, we denote by $e_{ab}$ the edge connecting vertices $a$ and $b$. Since the edges host spins, the total number of sites is
\begin{equation}
    N = \binom{M}{2}.
\end{equation}

For every vertex $a$, define the commuting and non-commuting operators, respectively,
\begin{align}
    P_a^{\mathrm C}
    &= \bigotimes_{b\neq a} X_{e_{ab}},\\
    P_a^{\mathrm{NC}}
    &= \left(\bigotimes_{b<a} Z_{e_{ab}}\right)
       \left(\bigotimes_{b>a} X_{e_{ab}}\right).
\end{align}
where $X,Y,Z$ denote Pauli matrices. The operators $P_a^{\mathrm C}$ mutually commute, whereas
\begin{equation}
    \{P_a^{\mathrm{NC}},P_b^{\mathrm{NC}}\}=0,
    \qquad a\neq b.
\end{equation}
We therefore define the commuting and non-commuting battery
Hamiltonians as
\begin{equation}
    H_B^{\mathrm C}=\sum_{a=1}^{M}P_a^{\mathrm C},
    \qquad
    H_B^{\mathrm{NC}}=\sum_{a=1}^{M}P_a^{\mathrm{NC}},
\end{equation}
respectively.

Next, for the charging, we identify a perfect matching of the vertices, which we may take to be
\begin{equation}
    \mathcal{M}
    =\{(1,2),(3,4),\ldots,(M-1,M)\}.
\end{equation}
We define
\begin{equation}
H_C
=
i\sum_{\substack{a=1\\ a\ {\rm odd}}}^{M-1}
Y_{e_{a,a+1}}\,\chi_C\gamma_{r(a)},
\end{equation}
where $\chi_C$ and $\gamma_r$, $r=1,\ldots,M/2$, are Majorana operators satisfying
\begin{equation}
\{\gamma_r,\gamma_s\}=2\delta_{rs}\mathbb{I},
\qquad
\{\chi_C,\gamma_r\}=0,
\qquad
\chi_C^2=\mathbb{I}.
\end{equation}
Because the $Y$ operators act on distinct matching edges and therefore commute, while the bilinears $i\chi_C\gamma_r$ mutually anticommute, the charging terms satisfy
\begin{equation}
\left\{
iY_{e_{a,a+1}}\chi_C\gamma_{r(a)},
iY_{e_{b,b+1}}\chi_C\gamma_{r(b)}
\right\}
=0,
\qquad a\neq b.
\end{equation}
Each charging term contains an even number of Majorana operators and is therefore fermion-parity even. The battery and charging geometries are schematically represented in Fig.~\ref{fig:K6_hamiltonian_geometry}.

With respect to the battery lattice, the charging is supported through one-local terms on the chosen perfect-matching edges and hence has effective locality
\begin{equation}
    k_C^{\mathrm{eff}}=1.
\end{equation}

Crucially, the commutator, the Hamiltonian operator norms, the $g$-extensivities and localities of $H_B$ and $H_C$ can be readily calculated from the algebra between their terms; we refer to the Supplemental Material \cite{SupplementalMaterial} for an explicit derivation. We obtain
\begin{align}
    \left\|[H_B^{\mathrm C},H_C]\right\|
        &=2\sqrt{2M},\\
    \left\|[H_B^{\mathrm{NC}},H_C]\right\|
        &=\sqrt{2}\,M,
\end{align}
while
\begin{equation}
    g_B^{\mathrm C}=g_B^{\mathrm{NC}}=2,
    \qquad
    k_C^{\mathrm{eff}}=1,
    \qquad
    \|H_C\|=\sqrt{\frac{M}{2}}.
\end{equation}
Consequently, in both cases,
\begin{equation}
    2k_C^{\mathrm{eff}}g_B\|H_C\|
    =2\sqrt{2M}.
\end{equation}
Thus, this bound captures the commutator norm exactly in the commuting case. In the non-commuting case, however, the commutator exceeds it by a factor
\begin{equation}
    \frac{
        \|[H_B^{\mathrm{NC}},H_C]\|
    }{
        2k_C^{\mathrm{eff}}g_B\|H_C\|
    }
    =
    \frac{\sqrt{M}}{2}.
\end{equation}

\begin{figure*}[t!]
    \centering

    \begin{overpic}[height=0.22\textheight]
        {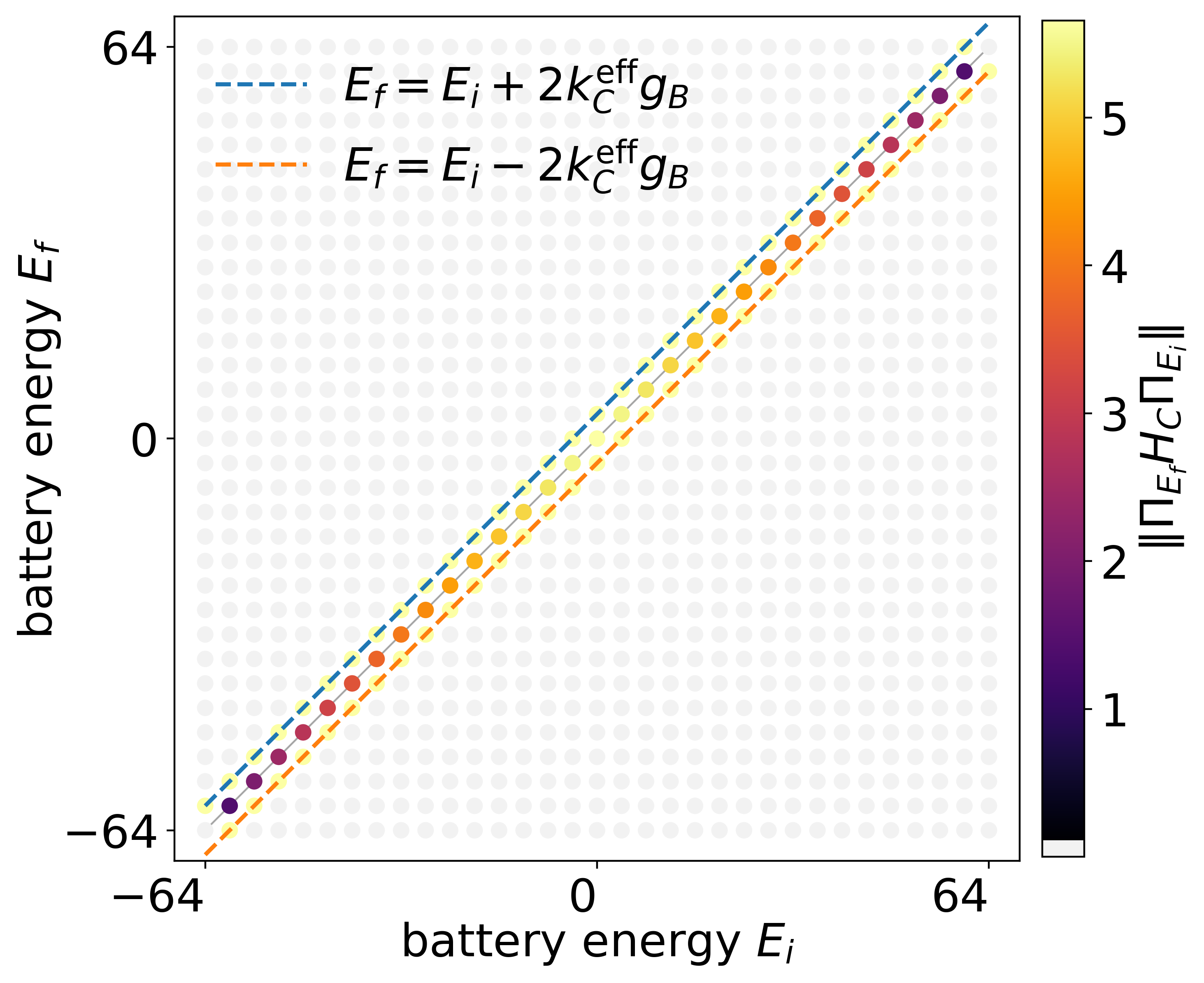}
        \put(3,92){\small\textbf{(a)}}
    \end{overpic}
    \hfill
    \begin{overpic}[height=0.22\textheight]
        {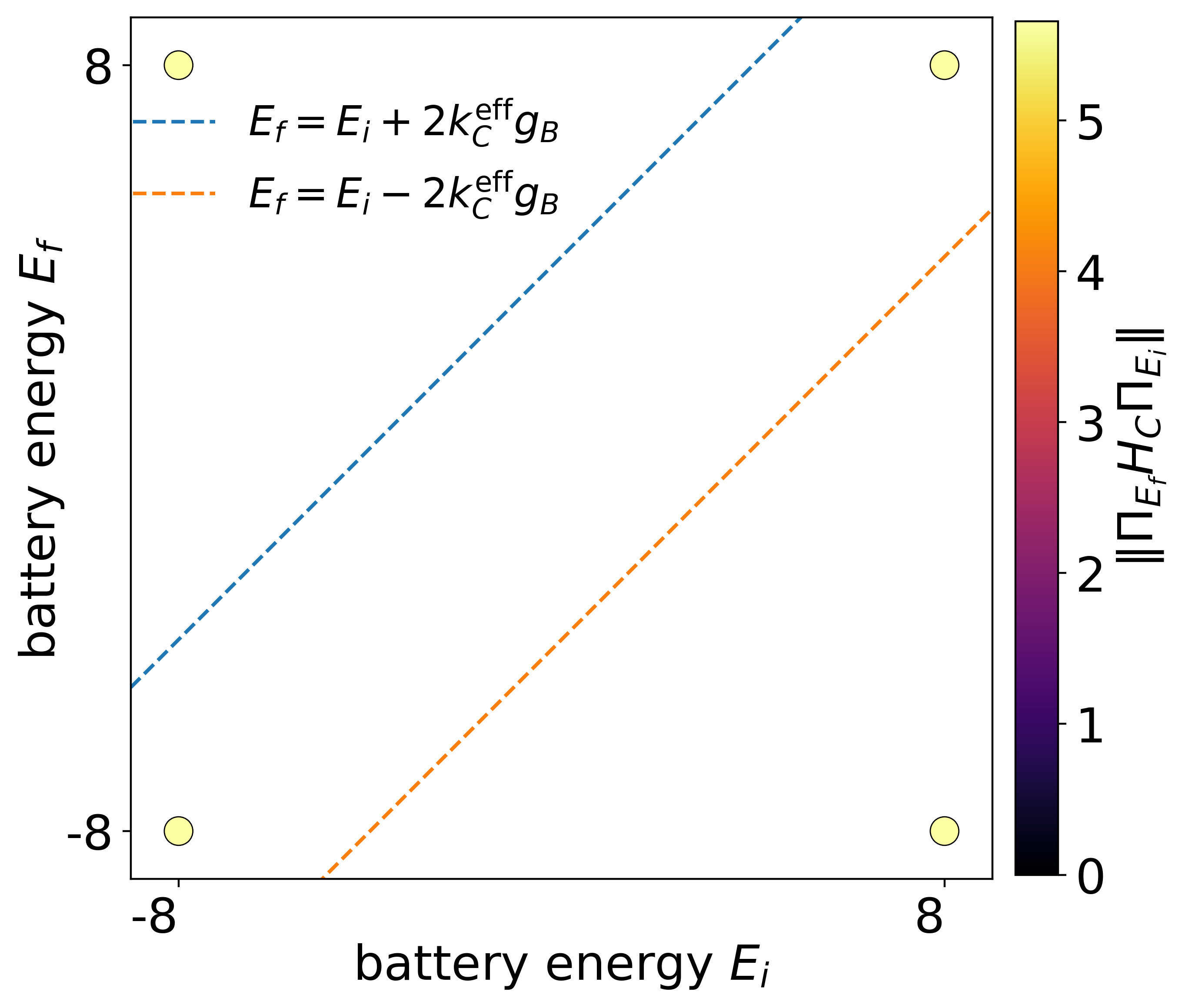}
        \put(3,92){\small\textbf{(b)}}
    \end{overpic}
    \hfill
    \begin{overpic}[height=0.22\textheight]
        {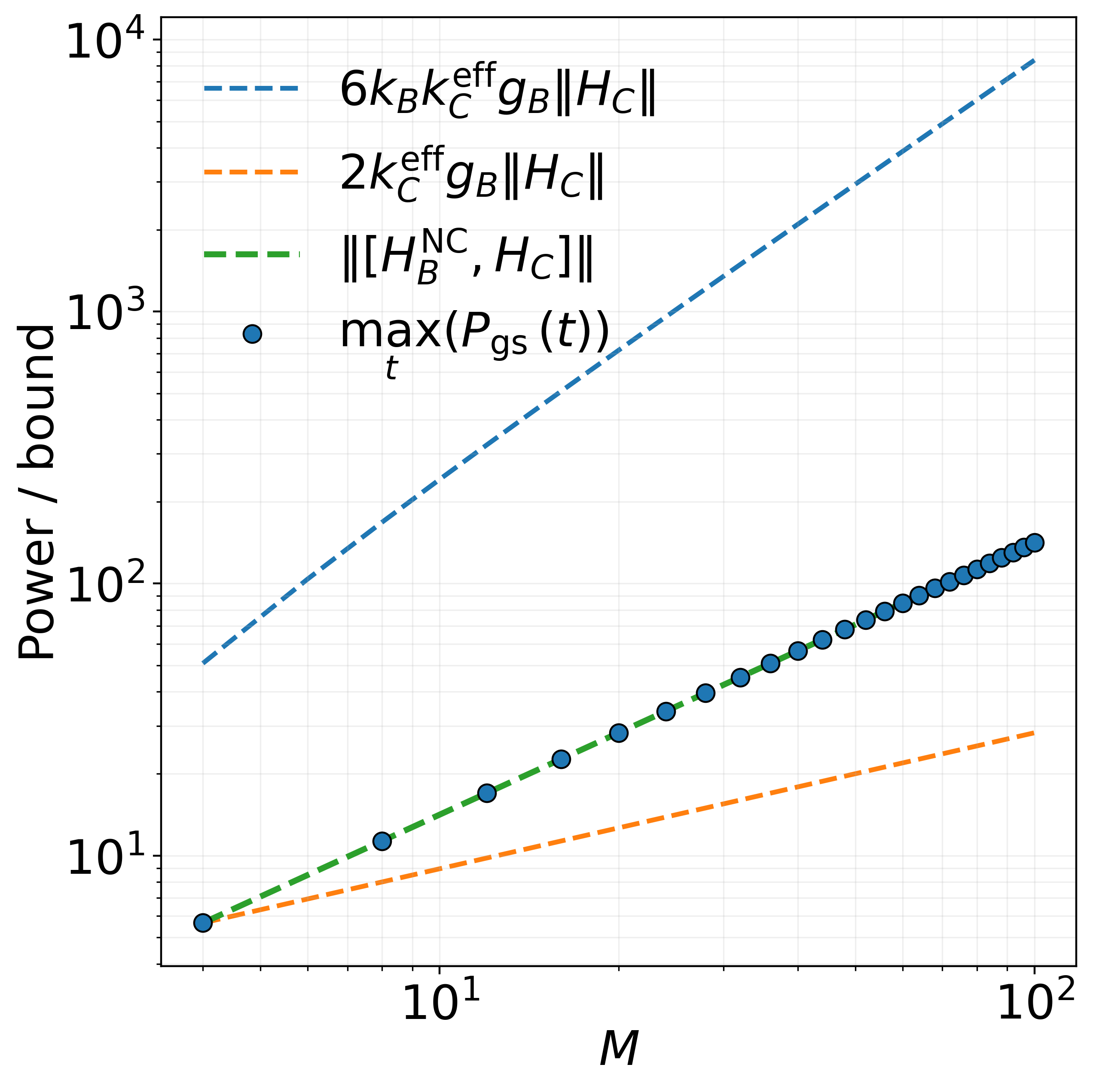}
        \put(3,106){\small\textbf{(c)}}
    \end{overpic}

    \caption{AKLH spectral transitions and dynamical consequences for the complete-graph construction. Panels (a) and (b) show the transition structure of the same charging acting on the commuting and noncommuting batteries, respectively, for $M=64$. In the commuting case, the support of $\|\Pi_{E_f}H_C\Pi_{E_i}\|$ is confined to the strip $|E_f-E_i|\leq 2k_C^{\mathrm{eff}}g_B$. In the noncommuting case, full-strength transitions occur between sectors separated by $|E_f-E_i|=2\sqrt{M}$. Panel (c) compares the analytical bounds, exact commutator norms, and maximal charging power attainable from suitable joint states supported in the battery ground-energy sector. All three panels are obtained from the corresponding exact analytical expressions.}
    \label{fig:aklh_and_dynamics}
\end{figure*}

On the other hand, noting that
\begin{equation}
    k_B=M-1,
\end{equation}
the general non-commuting bound gives
\begin{equation}
    6k_Bk_C^{\mathrm{eff}}g_B\|H_C\|
    =
    6(M-1)\sqrt{2M},
\end{equation}
which remains valid, while the upper bound applicable to commuting batteries is exceeded in the non-commuting construction. Fig.~\ref{fig:aklh_and_dynamics}(c) illustrates this by plotting the exact analytical scalings of the relevant quantities.

Starting from a suitably chosen state of the extended system, supported entirely in the ground-energy sector of the noncommuting battery, we can establish that an instantaneous power saturating the commutator norm is achieved. The relevant joint ground sector is highly degenerate, and the choice of joint battery–auxiliary vector influences the attainable power. The saturating vector need not be a product across the battery and auxiliary degrees of freedom. We refer to the Supplemental Material \cite{SupplementalMaterial} for its construction and for closed-form expressions for the resulting evolution and power.

Finally, we note that the Hamiltonians can be rescaled to be energy-extensive. Including the $O(M)$ fermionic degrees of freedom, the total system size remains
$N_{\rm tot}=N+O(M)=\Theta(N)$, with $N=\binom{M}{2}=\Theta(M^2)$, so all asymptotic scalings in $N$ are unchanged. We make the transformations
\begin{align*}
H_B^{\mathrm C}
    &\rightarrow \frac{N}{M}H_B^{\mathrm C},
&
H_B^{\mathrm{NC}}
    &\rightarrow \frac{N}{\sqrt M}H_B^{\mathrm{NC}},
\\
H_C
    &\rightarrow \sqrt{\frac{2}{M}}\,N H_C.
\end{align*}
With these rescalings,
\begin{equation}
    \|H_B^{\mathrm C}\|
    =
    \|H_B^{\mathrm{NC}}\|
    =
    \|H_C\|
    =N.
\end{equation}

The corresponding commutator norms then scale as
\begin{align}
    \left\|
        [H_B^{\mathrm C},H_C]
    \right\|
    &\sim N^{3/2},\\
    \left\|
        [H_B^{\mathrm{NC}},H_C]
    \right\|
    &\sim N^2.
\end{align}

Although the extensive normalization correspondingly rescales \(g_B\), the dimensionless ratio \(\|[H_B,H_C]\|/(2k_C^{\mathrm{eff}}g_B\|H_C\|)\) is invariant under independent rescalings of \(H_B\) and \(H_C\); hence, the parametrically growing violation found above is unchanged. Therefore, non-commutativity here enables the quadratic superextensive scaling of the charging power.

\prlsection{Discussion} The model exposes a stark distinction between the transition structure of commuting and non-commuting batteries. For a commuting battery, with $g_i^B$ being the $g$-extensivity of site $i$,
\begin{equation}
    g_i^B=\sum_{X\ni i}\|h_X^B\|,
\end{equation}
a charging term $h_Y^C$ can only change the mutually commuting battery-energy contributions $h_X^B$ that overlap its support, and the accessible energy transitions are controlled more precisely by
\begin{equation}
    G_B(Y)=\sum_{i\in Y}g_i^B.
\end{equation}
In particular,
\begin{equation}
    \Pi_{E'}^B h_Y^C \Pi_E^B=0,
    \qquad
    \text{for}\qquad
    |E'-E|>2G_B(Y).
\end{equation}
Since
\begin{equation}
    G_B(Y)\leq g_B k_C^{\mathrm{eff}},
\end{equation}
where $k_C^{\mathrm{eff}}$ counts only the locality of the charging term acting non-trivially on the battery support, we obtain an upper bound of
\begin{equation}
    2k_C^{\mathrm{eff}}g_B\|H_C\|
\end{equation}
on the maximum instantaneous power.

In our $H_B^{\mathrm{NC/C}}$ model on the complete graph, the only allowed transitions in the commuting case are
\begin{equation}
    \Delta E=0,\pm4,
\end{equation}
precisely saturating
\begin{equation}
    2g_Bk_C^{\mathrm{eff}}=4.
\end{equation}

In contrast, for the anticommuting battery,
\begin{equation}
    \left(H_B^{\mathrm{NC}}\right)^2=M\mathbb{I},
\end{equation}
and the charging has a non-zero matrix element between its two energy sectors,
\begin{equation}
    E_\pm=\pm\sqrt{M},
    \qquad
    \left\|\Pi_+^B H_C\Pi_-^B\right\|
    =\sqrt{\frac{M}{2}},
\end{equation}
with energy separation
\begin{equation}
    |E_+-E_-|=2\sqrt{M}.
\end{equation}
This is shown in the AKLH transition plots of Fig.~\ref{fig:aklh_and_dynamics}(a) and Fig.~\ref{fig:aklh_and_dynamics}(b), and a proof is given in the Supplemental Material \cite{SupplementalMaterial}. The transition plots reveal the spectral origins for enhancement of power in the non-commuting case.

\prlsection{Conclusion} We have shown that power bounds valid for commuting batteries need not survive in non-commuting systems. For commuting $H_B$, the accessible energy transitions for charging terms are controlled by the $g$-extensivities of the battery sites they actually touch. For non-commuting $H_B$, the bound becomes dependent on the battery locality $k_B$, which can lead to a better scaling for the maximum power.

These results also suggest several immediate extensions. Our construction shows that battery locality influences the bound in spin systems; however, the non-commuting $k_B$-dependent bound is not saturated by our model. Finding a family of Hamiltonians that saturate it would establish tightness.

Finally, the power depends on the choice of joint battery–auxiliary vector within the battery ground-energy sector. Whether the same enhancement can be attained from an initially uncorrelated battery–auxiliary state remains open. Understanding robustness, non-degenerate batteries, finite-temperature initial states, and microscopic realizations in experimental architectures remain natural next steps.

Broadly, we conclude that the battery Hamiltonian is not merely an energy observable: its operator algebra can actively enable transitions forbidden in commuting systems.

\prlsection{Acknowledgments}
The author gratefully acknowledges Dario Rosa, Anupam Sarkar, and Sibasish Ghosh for fruitful discussions.  The author also acknowledges the use of GPT 5.6 for language refinement and improving the presentation of the manuscript.

\bibliographystyle{apsrev4-2}
\bibliography{bibliography}

\onecolumngrid
\appendix
\section*{Supplemental Material}

\section*{Locality bounds as a graph coloring problem}
\label{sec:SM_coloring}

In this section, we review the origin of the battery-locality factor in the non-commuting power bound and reformulate it as a coloring problem. This viewpoint will also motivate the complete-graph construction used in the main text. A complete proof of the non-commuting bound appeared first in \cite{Kuwahara2016} and was adapted to the quantum battery context in \cite{SarkarGhosh2025}; we provide a derivation of their rigorous results here for completeness. 

Consider a $k$-local, $g$-extensive Hamiltonian
\begin{equation}
    H=\sum_{X\subseteq\Lambda} h_X,
    \qquad |X|\leq k,
\end{equation}
with
\begin{equation}
    \sum_{X\ni i}\|h_X\|\leq g
    \qquad
    \forall\,i\in\Lambda.
    \label{eq:g_extensive_SM}
\end{equation}
For a general $q$-local operator $\Gamma^{(q)}$, the bound of \cite{Kuwahara2016} gives
\begin{equation}
    \left\|[H,\Gamma^{(q)}]\right\|
    \leq
    6gkq\,\|\Gamma^{(q)}\|.
    \label{eq:Kuwahara_general_SM}
\end{equation}
The additional factor $k$, absent when $H$ itself is commuting, can be understood from the decomposition of $H$ into commuting pieces \cite{Kuwahara2016}.

\subsection*{Commuting decomposition as a coloring problem}

To make this structure explicit, we first discretize the strengths of the interaction terms.  For a small $\epsilon>0$, define $  \widetilde h_X = \epsilon\,\frac{h_X}{\|h_X\|},$ and use $N_X = \left\lfloor\frac{\|h_X\|}{\epsilon} \right\rfloor$ copies of $\widetilde h_X$ to define a discretized Hamiltonian:
\begin{equation}
    H^{(\epsilon)}
    =
    \sum_X N_X\widetilde h_X,
    \label{eq:discretized_H}
\end{equation}
It then satisfies $\|H-H^{(\epsilon)}\|=O(\epsilon N).$ The $g$-extensivity condition implies that the number of such $\epsilon$-strength interaction copies incident on any site further satisfies $\sum_{X\ni i}N_X \leq \frac{g}{\epsilon}.$

We may now associate a \emph{conflict graph} $\mathcal C_H$ with the interaction terms of $H^{(\epsilon)}$, considering each of the $N_X$ copies of each $\widetilde h_X$ distinctly.  Each vertex of $\mathcal C_H$ represents one interaction term $\widetilde h_X$, and two vertices are connected whenever their supports overlap, $X\cap X'\neq\varnothing.$
A proper coloring of this graph partitions the Hamiltonian into color classes such that no two terms of a given color overlap.  Thus, if $\mathcal I_m$ denotes the terms assigned color $m$, then
\begin{equation}
    X\cap X'=\varnothing
    \qquad \forall \
    h_{X},h_{X'}\in\mathcal I_m,
\end{equation}
and consequently $ H_m^{\rm c} = \sum_{h_X\in\mathcal I_m}\widetilde h_X$ is a commuting Hamiltonian, $[\widetilde h_X,\widetilde h_{X'}]=0 \ \forall \ h_{X},h_{X'}\in\mathcal I_m.$

The number of colors required is depends on the locality $k$. Indeed, a term supported on $X$, with $|X|\leq k$, can conflict with terms incident on each of its at most $k$ sites.  Using $\sum_{X\ni i}N_X \leq \frac{g}{\epsilon}$, its degree in the conflict graph is upper-bounded parametrically by $k(\frac{g}{\epsilon} -1)$. Accordingly, the interaction terms can be organized into
\begin{equation}
    \bar n
    =
    O\!\left(\frac{kg}{\epsilon}\right)
    \label{eq:number_colors}
\end{equation}
commuting color classes (since a graph of degree $k(\frac{g}{\epsilon} -1)$ needs at most $k(\frac{g}{\epsilon} -1) + 1$ colors). The construction of \cite{Kuwahara2016} gives, more specifically,
\begin{equation}
    \bar n
    =
    k\left\lfloor\frac{g}{\epsilon}\right\rfloor,
\end{equation}
and permits the decomposition
\begin{equation}
    H^{(\epsilon)}
    =
    \sum_{m=1}^{\bar n} H_m^{\rm c}
    =
    \frac{1}{\bar n}
    \sum_{m=1}^{\bar n}
    \overline H_m^{\rm c},
    \qquad
    \overline H_m^{\rm c}
    :=
    \bar n H_m^{\rm c}.
    \label{eq:commuting_decomposition}
\end{equation}
Each $\overline H_m^{\rm c}$ is commuting, remains $k$-local, and is at most $(gk)$-extensive.  Taking $\epsilon\rightarrow0$ then yields the corresponding decomposition of $H$.

This explains the locality factor in Eq.~\eqref{eq:Kuwahara_general_SM}.  For a commuting Hamiltonian $H^{\rm c}$ with extensivity $g$, \cite{Kuwahara2016} first establishes
\begin{equation}
    \left\|[H^{\rm c},\Gamma^{(q)}]\right\|
    \leq
    6gq\,\|\Gamma^{(q)}\|.
    \label{eq:commuting_kuwahara}
\end{equation}
They then apply this result to every $\overline H_m^{\rm c}$ in Eq.~\eqref{eq:commuting_decomposition}, whose extensivity is bounded by $gk$, to obtain:
\begin{align}
    \left\|[H,\Gamma^{(q)}]\right\|
    &=
    \left\|
    \frac{1}{\bar n}
    \sum_{m=1}^{\bar n}
    [\overline H_m^{\rm c},\Gamma^{(q)}]
    \right\|
    \nonumber\\
    &\leq
    \frac{1}{\bar n}
    \sum_{m=1}^{\bar n}
    \left\|
    [\overline H_m^{\rm c},\Gamma^{(q)}]
    \right\|
    \nonumber\\
    &\leq
    6gkq\,\|\Gamma^{(q)}\|,
\end{align}
which reproduces Eq.~\eqref{eq:Kuwahara_general_SM}. Thus, the factor $k$ may be viewed as the cost of resolving a noncommuting Hamiltonian into commuting color classes.

\subsection*{Refinement to the effective charging support}

For the quantum-battery application, the locality of the charging Hamiltonian can be sharpened because degrees of freedom on which a commuting battery Hamiltonian does not act cannot contribute to battery-energy transitions. Let $H_B=\sum_X h_X^B$ be a commuting battery Hamiltonian and consider a single charging term $h_Y^C$ supported on $Y$. Let $H_B=H_{B,Y}+H_{B,\overline{Y}}$, where $H_{B,Y}=\sum_{X:X\cap Y\neq\varnothing}h_X^B$ and $H_{B,\overline{Y}} =\sum_{X:X\cap Y=\varnothing}h_X^B$. Since all battery terms commute, $[H_{B,Y},H_{B,\overline Y}]=0,$ while locality implies $[h_Y^C,H_{B,\overline Y}]=0.$
Therefore $h_Y^C$ cannot change the eigenvalue of $H_{B,\overline Y}$, and any battery-energy transition generated by $h_Y^C$ must arise entirely from $H_{B,Y}$. Hence
\begin{equation}
    \Pi_{E'}^B h_Y^C\Pi_E^B=0
    \qquad
    {\rm if}
    \qquad
    |E'-E|>2\|H_{B,Y}\|.
    \label{eq:exact_active_transition}
\end{equation}

Define the local battery load $g_i^B:=\sum_{X\ni i}\|h_X^B\|,$ and the battery-active load of the support $Y$ as
\begin{equation}
    \mathcal G_B(Y)
    :=
    \sum_{i\in Y}g_i^B.
    \label{eq:GBY}
\end{equation}
By the triangle inequality,
\begin{align}
    \|H_{B,Y}\|
    \leq
    \sum_{X:X\cap Y\neq\varnothing}\|h_X^B\|
    \leq
    \sum_{i\in Y}\sum_{X\ni i}\|h_X^B\|
    =
    \mathcal G_B(Y).
\end{align}
Consequently,
\begin{equation}
    \boxed{
    \Pi_{E'}^B h_Y^C\Pi_E^B=0
    \quad {\rm for}\quad
    |E'-E|>2\mathcal G_B(Y).
    }
    \label{eq:GB_transition}
\end{equation}

For $H_C=\sum_Y h_Y^C$, further define $\mathcal G_{B|C} := \max_{Y:h_Y^C\neq0}\mathcal G_B(Y).$
Since every term in $H_C$ separately obeys
Eq.~\eqref{eq:GB_transition}, linearity gives
\begin{equation}
    \boxed{
    \Pi_{E'}^B H_C\Pi_E^B=0
    \quad {\rm for}\quad
    |E'-E|>2\mathcal G_{B|C}.
    }
    \label{eq:full_HC_transition}
\end{equation}

For a battery with $g_i^B\leq g_B,$ we finally only need an effective charging locality
\begin{equation}
    k_C^{\rm eff}
    :=
    \max_{Y:h_Y^C\neq0}
    \left|
    \left\{
    i\in Y:g_i^B>0
    \right\}
    \right|,
\end{equation}
so that $\mathcal G_{B|C}\leq g_B k_C^{\rm eff}.$ The conventional scale $g_Bk_C$ is therefore recovered as a looser bound, while degrees of freedom belonging only to the charger need not be counted.

Replacing the transition width $2gq$ in the commuting proof by $2\mathcal G_{B|C}$ yields
\begin{equation}
    \boxed{
    \|[H_B,H_C]\|
    \leq
    6\mathcal G_{B|C}\|H_C\|
    \leq
    6g_Bk_C^{\rm eff}\|H_C\|.
    }
    \label{eq:refined_commuting_bound}
\end{equation}
Applying the commuting decomposition discussed above to a non-commuting $k_B$-local battery gives correspondingly
\begin{equation}
    \boxed{
    \|[H_B,H_C]\|
    \leq
    6k_B\mathcal G_{B|C}\|H_C\|
    \leq
    6k_Bg_Bk_C^{\rm eff}\|H_C\|.
    }
    \label{eq:refined_noncommuting_bound}
\end{equation}
The numerical prefactor is not central to the present work, since the parametric asymptotic scaling of power is independent of it; the important distinction is the appearance of the battery-locality factor $k_B$ only in the noncommuting estimate.

\subsection*{Why the complete-graph battery is a natural candidate}

The vertex coloring picture gives a strategy for constructing a suitable model in which the battery-locality dependence can become indispensable: we need a $k_B$-local battery whose interaction terms themselves require $\Theta(k_B)$ distinct colors. Our noncommuting complete-graph battery has precisely this property.

Let the physical spins occupy the edges of $K_M$, and associate one battery term with every vertex, $H_B^{\rm NC} = \sum_{a=1}^{M}P_a^{\rm NC},$ where $ P_a^{\rm NC}=\left(\bigotimes_{b<a}Z_{e_{ba}}\right)\left(\bigotimes_{b>a}X_{e_{ab}}\right).$ It is supported on the star $S_a = \{e_{ab}:b\neq a\},$ so that $|S_a|=M-1 \Longrightarrow k_B=M-1.$ For any two distinct vertices $a$ and $b$, $S_a\cap S_b=\{e_{ab}\}.$ Moreover, on this unique common spin, one of the two strings acts with $X$ and the other with $Z$.  Hence $\{P_a^{\rm NC},P_b^{\rm NC}\}=0,\ a\neq b.$
The conflict graph of the $M$ battery terms is therefore itself the complete graph, $\mathcal C_{H_B^{\rm NC}}=K_M.$ 

We need at most $\chi(\mathcal C_{H_B^{\rm NC}}) = \chi(K_M) = M$ colors. Indeed, since no two distinct $P_a^{\rm NC}$ can belong to the same commuting component, any partition of the battery terms into mutually commuting classes requires also at least $M$ classes. Thus the complete-graph construction realizes the maximal coloring cost suggested by the locality-dependent non-commuting bound: dividing the noncommuting battery into commuting pieces requires a number of components proportional to its own locality.

This is the principal reason that $H_B^{\rm NC}$ is a natural candidate for testing whether the $k_B$ appearing in the generic bound can become operational.

\section*{Algebra and spectral properties of the complete-graph batteries}
\label{sec:SM_battery_algebra}

We now derive the basic algebraic properties of the commuting and
noncommuting battery Hamiltonians introduced in the main text.
\subsection*{Commuting battery}

The commuting star operators are $P_a^{\rm C} := \bigotimes_{b\neq a}X_{e_{ab}},$ they have eigenvalues $\pm 1$, and the corresponding battery Hamiltonian is
\begin{equation}
    H_B^{\rm C}
    =
    \sum_{a=1}^{M}P_a^{\rm C}.
    \label{eq:HBC_def_SM}
\end{equation}
For two distinct vertices $a\neq b$, the stars $S_a$ and $S_b$
intersect on exactly one edge, $e_{ab}$. On this common spin both operators act with $X$, while their remaining supports are disjoint.  Therefore
\begin{equation}
    [P_a^{\rm C},P_b^{\rm C}]=0
    \qquad
    \forall\,a,b.
    \label{eq:PC_commute}
\end{equation}
There is one global relation among the $M$ stars.  Since every edge $e_{ab}$ belongs to precisely the two stars $S_a$ and $S_b$, $ \prod_{a=1}^{M}P_a^{\rm C} = \prod_{1\leq a<b\leq M}X_{e_{ab}}^2 = I.$ Thus a simultaneous eigenstate of all $P_a^{\rm C}$ may be labelled by eigenvalues $p_a=\pm1,$ subject to $\prod_{a=1}^{M}p_a=1.$
Equivalently, the number $n_-$ of negative star eigenvalues must be even. The corresponding battery energy is
\begin{equation}
    E
    =
    \sum_{a=1}^{M}p_a
    =
    M-2n_-.
\end{equation}
Since $n_-=0,2,\ldots,M$ for even $M$, the spectrum is $ E^{\rm C} = -M,-M+4,\ldots,M-4,M.$ In particular, the simultaneous $p_a=+1$ sector gives $\|H_B^{\rm C}\|=M.$ The $M$ commuting stars have one independent relation, so there are $M-1$ independent stabilizer generators.  Each allowed assignment $\{p_a\}$ therefore labels a joint eigenspace of dimension
\begin{equation}
    2^{N-(M-1)}
    =
    2^{N-M+1}.
    \label{eq:HBC_sector_deg}
\end{equation}
This large degeneracy will show up when discussing the projected transition structure below.

\subsection*{Noncommuting battery}

The noncommuting star operators are defined as $ P_a^{\rm NC}   :=   \left(  \bigotimes_{b<a}Z_{e_{ba}}  \right)  \left( \bigotimes_{b>a}X_{e_{ab}} \right)$, have eigenvalues $\pm 1$ with
\begin{equation}
    H_B^{\rm NC}
    =
    \sum_{a=1}^{M}P_a^{\rm NC}.
    \label{eq:HBNC_def_SM}
\end{equation}
Consider two vertices $a<b$.  Their stars intersect only at
$e_{ab}$. Now,
$P_a^{\rm NC}$ acts with $X_{e_{ab}}$, whereas
$P_b^{\rm NC}$ acts with $Z_{e_{ab}}$. The $M$ star operators therefore pairwise anticommute, and furnish a Clifford algebra.

This immediately determines the square of the battery Hamiltonian:
\begin{align}
    (H_B^{\rm NC})^2 =
    \sum_{a=1}^{M}(P_a^{\rm NC})^2
    +
    \sum_{a<b}
    \{P_a^{\rm NC},P_b^{\rm NC}\}
    =
    MI.
    \label{eq:HBNC_square}
\end{align}
Consequently every eigenvalue $E$ of $H_B^{\rm NC}$ satisfies $ E^2=M.$
The spectrum therefore has only two distinct highly-degenerate energy eigenvalues, $E_\pm^{\rm NC}=\pm\sqrt{M}, $
and $\|H_B^{\rm NC}\|=\sqrt M.$

Since each nontrivial Pauli string $P_a^{\rm NC}$ is traceless, Tr$(H_B^{\rm NC}=0).$
Together with the two-level spectrum, this implies equal multiplicities of the two eigenvalues.  Writing $\Pi_\pm$ for the corresponding projectors, $H_B^{\rm NC} = \sqrt M(\Pi_+-\Pi_-)$, with rank$(\Pi_+)=$ rank$(\Pi_-)=2^{N-1}$. The ground space of $H_B^{\rm NC}$ is therefore highly degenerate, and we will use this when considering charging dynamics.

\subsection*{Locality and $g$-extensivity}

The two batteries have identical support geometry.  Every term $P_a^{\rm C}$ or $P_a^{\rm NC}$ is supported on the full star $S_a$, which contains $M-1$ edge spins.  Hence $k_B^{\rm C} = k_B^{\rm NC} = M-1.$ The local interaction strength is also identical.  A physical edge spin $e_{ab}$ occurs in precisely two battery terms: the star associated with vertex $a$ and that associated with vertex $b$.  Since both terms have unit operator norm, $g_B^{\rm C}= g_B^{\rm NC}=2.$

\section*{Charging Hamiltonian and its algebra}
\label{sec:SM_charger_algebra}

We now derive the properties of the charging Hamiltonian. The same charging is used for both the commuting and noncommuting batteries. Throughout, $M$ is even and we write $m:=\frac{M}{2}$. We choose the perfect matching $\mathcal M=\{(1,2),(3,4),\ldots,(M-1,M)\}.$
Equivalently, the $r$th matching edge is
\begin{equation}
    e_r=e_{2r-1,2r},
    \qquad
    r=1,\ldots,m.
\end{equation}

To each matching edge we associate an auxiliary Majorana mode $\gamma_r$, together with one additional Majorana mode $\chi_C$. They satisfy $  \chi_C^\dagger=\chi_C,  \ \gamma_r^\dagger=\gamma_r,  \ \chi_C^2=\gamma_r^2=I, \ \{\gamma_r,\gamma_s\}=2\delta_{rs}I,\ \{\chi_C,\gamma_r\}=0.$ The charging Hamiltonian is
\begin{equation}
    H_C
    =
    i\sum_{\substack{a=1\\a\ {\rm odd}}}^{M-1}
    Y_{e_{a,a+1}}\,
    \chi_C\gamma_{r(a)},
    \qquad
    r(a)=\frac{a+1}{2}.
    \label{eq:HC_def_SM}
\end{equation}
Equivalently, $ H_C = \sum_{r=1}^{m}Q_r, \ Q_r:=iY_{e_r}\chi_C\gamma_r.$ The operators $\chi_C$ and $\gamma_r$ act only on the auxiliary fermionic degrees of freedom, while $Y_{e_r}$ acts on the battery edge qubit $e_r$. Hence
\begin{equation}
    [Y_{e_r},\chi_C]
    =
    [Y_{e_r},\gamma_s]
    =
    0
    \qquad
    \forall\,r,s.
    \label{eq:Y_Majorana_commute}
\end{equation}
The Majorana factor $i\chi_C\gamma_r$ is Hermitian and fermion-parity even, so that each $Q_r$ is a Hermitian charging term.

\subsection*{Mutual anticommutation of the charger terms}

We first note that $Q_r^\dagger = Q_r,$ and $ Q_r^2 = I.$ For two distinct matching edges $r\neq s$, the corresponding Pauli operators act on different edge qubits and therefore commute, $ [Y_{e_r},Y_{e_s}]=0.$ Using the Majorana algebra, $ Q_rQ_s = Y_{e_r}Y_{e_s}\gamma_r\gamma_s = - Q_sQ_r.$ Therefore
\begin{equation}
    \boxed{
    \{Q_r,Q_s\}=0,
    \qquad
    r\neq s.
    }
    \label{eq:Qr_anticommute}
\end{equation}
Thus the $m=M/2$ charging terms form a mutually anticommuting family. It follows immediately that
\begin{align}
    H_C^2 =
    \sum_{r=1}^{m}Q_r^2
    +
    \sum_{r<s}\{Q_r,Q_s\}
    =
    mI
    =
    \frac{M}{2}I.
    \label{eq:HC_square}
\end{align}
Hence the charger has only the two eigenvalues, $ \pm\sqrt{\frac{M}{2}}$, and therefore
\begin{equation}
    \boxed{
    \|H_C\|
    =
    \sqrt{\frac{M}{2}}.
    }
    \label{eq:HC_norm_SM}
\end{equation}

\subsection*{Effective locality of the charger}

Each charging term $Q_r$ acts on one battery degree of freedom, namely the matching-edge spin $e_r$, together with auxiliary fermionic degrees of freedom that do not appear in the battery Hamiltonian. Consequently, using the battery-active load introduced earlier, the support $Y_r$ of the $r$th charger term therefore has $\mathcal G_B(Y_r) = \sum_{i\in Y_r}g_i^B = 2.$
    
Equivalently, each charging term acts nontrivially on only one degree of freedom on which the battery Hamiltonian itself has support. Hence the effective charger locality is $ k_C^{\rm eff}=1.$ Since $g_B=2$ for both battery Hamiltonians, $g_Bk_C^{\rm eff}=2. $ For the commuting battery, the transition bound therefore becomes
\begin{equation}
    \Pi_{E'}^B H_C\Pi_E^B=0
    \qquad
    {\rm for}
    \qquad
    |E'-E|>4.
    \label{eq:model_commuting_transition_width}
\end{equation}

\subsection{Useful algebra with the battery stars}

Let $Q_r=iY_{e_r}\chi_C\gamma_r$, then $ \{P_{2r-1},Q_r\}=  \{P_{2r},Q_r\}  =0, $ whereas $ [P_c,Q_r]=0, $ for $c\neq2r-1,2r.$ Thus every charging term anticommutes with exactly two battery interaction terms, namely the stars associated with the vertices of its matching edge. The local battery-charger overlap pattern is therefore identical for the commuting and noncommuting batteries. The difference between the two cases comes entirely from the mutual algebra of the battery terms themselves. As shown below, this determines whether the contributions from different matching pairs combine in quadrature or coherently in the commutator.

\section*{Norm of the Commutator}
\label{sec:SM_NC_commutator}
For the commuting and non-commuting complete-graph batteries,
\begin{equation}
    H_B^{\mathrm{C/NC}}
    =
    \sum_{a=1}^{M} P_a^{\mathrm{C/NC}},
    \qquad
    H_C
    =
    \sum_{r=1}^{M/2} Q_r,
\end{equation}
we have
\begin{equation}
    \left[P_a^{\mathrm{C/NC}},Q_r\right]
    =
    \begin{cases}
        2P_a^{\mathrm{C/NC}}Q_r,
        & a=2r-1 \ \text{or}\ 2r,\\[4pt]
        0,
        & \text{otherwise}.
    \end{cases}
\end{equation}
Consequently,
\begin{align}
    \left[H_B^{\mathrm{C/NC}},H_C\right]
    &=
    \sum_{a=1}^{M}
    \sum_{r=1}^{M/2}
    \left[P_a^{\mathrm{C/NC}},Q_r\right]
    \\
    &=
    \sum_{r=1}^{M/2}
    2\left(
        P_{2r-1}^{\mathrm{C/NC}}
        +
        P_{2r}^{\mathrm{C/NC}}
    \right)Q_r.
\end{align}

Now define, for the commuting and non-commuting cases,
\begin{equation}
    A_r^{\mathrm C}
    =
    2\left(
        P_{2r-1}^{\mathrm C}
        +
        P_{2r}^{\mathrm C}
    \right)Q_r,
    \qquad
    A_r^{\mathrm{NC}}
    =
    2\left(
        P_{2r-1}^{\mathrm{NC}}
        +
        P_{2r}^{\mathrm{NC}}
    \right)Q_r.
\end{equation}
For distinct matching pairs, one finds
\begin{equation}
    \left\{
        A_r^{\mathrm C},
        A_{r'}^{\mathrm C}
    \right\}
    =0,
    \qquad
    \left[
        A_r^{\mathrm{NC}},
        A_{r'}^{\mathrm{NC}}
    \right]
    =0,
    \qquad
    r\neq r'.
\end{equation}

For the commuting case, $ \left(A_r^{\mathrm C}\right)^2    =  -4\left(     P_{2r-1}^{\mathrm C}     +   P_{2r}^{\mathrm C}   \right)^2    =  -8\left(    \mathbb I      +     P_{2r-1}^{\mathrm C}P_{2r}^{\mathrm C} \right),$ where we used
\(
[P_{2r-1}^{\mathrm C},P_{2r}^{\mathrm C}]=0
\).
Since the $A_r^{\mathrm C}$ mutually anticommute,
\begin{align}
    -\left[H_B^{\mathrm C},H_C\right]^2
    =
    -\left(
        \sum_{r=1}^{M/2}A_r^{\mathrm C}
    \right)^2
    =
    -\sum_{r=1}^{M/2}
    \left(A_r^{\mathrm C}\right)^2 =
    8\sum_{r=1}^{M/2}
    \left(
        \mathbb I
        +
        P_{2r-1}^{\mathrm C}P_{2r}^{\mathrm C}
    \right).
\end{align}
The operators $P_a^{\mathrm C}$ mutually commute and possess a
simultaneous $+1$ eigenstate; explicitly, the state with every edge
spin in the $+1$ eigenstate of $X$ satisfies $  P_a^{\mathrm C}|\phi\rangle = |\phi\rangle, \  a=1,\ldots,M.$ Hence $ P_{2r-1}^{\mathrm C}P_{2r}^{\mathrm C}|\phi\rangle=|\phi\rangle$ for every $r$, and therefore
\begin{equation}
    \left\|
        -\left[H_B^{\mathrm C},H_C\right]^2
    \right\|
    =
    8M.
\end{equation}
Thus,
\begin{equation}
    \boxed{
    \left\|
        \left[H_B^{\mathrm C},H_C\right]
    \right\|
    =
    2\sqrt{2M}.
    }
\end{equation}

For the non-commuting case, $  \left(A_r^{\mathrm{NC}}\right)^2 =  -4\left(      P_{2r-1}^{\mathrm{NC}}      +   P_{2r}^{\mathrm{NC}}  \right)^2=-8\,\mathbb I,$ where we used $\left\{  P_{2r-1}^{\mathrm{NC}},  P_{2r}^{\mathrm{NC}}  \right\}   =0.$ Thus, $\left\|A_r^{\mathrm{NC}}\right\|=2\sqrt{2}.$ Since the $A_r^{\mathrm{NC}}$ mutually commute, define $ S_r  =  \frac{A_r^{\mathrm{NC}}}{2\sqrt{2}\,i},   \   S_r^2=\mathbb I,  \  [S_r,S_{r'}]=0.$ For completeness, the existence of a simultaneous $+1$ eigenstate of all the $S_r$ can be seen directly. Define $F_r =  iP_{2r-1}^{\mathrm{NC}}P_{2r}^{\mathrm{NC}}.$ These operators satisfy $ F_r^2=\mathbb I,   \  \{F_r,S_r\}=0,  \ [F_r,S_{r'}]=0 \quad (r'\neq r).$ Starting from any simultaneous eigenstate of the mutually commuting $S_r$, application of $F_r$ therefore flips the eigenvalue of $S_r$ without changing any of the other eigenvalues. Hence all joint eigenvalue sectors occur, and in particular there exists a state $|\psi\rangle$ such that $ S_r|\psi\rangle = |\psi\rangle \ \forall r.$
Equivalently,
\begin{equation}
    A_r^{\mathrm{NC}}|\psi\rangle
    =
    2\sqrt{2}\,i\,|\psi\rangle
    \qquad
    \forall r.
\end{equation}
It follows that
\begin{align}
    \left|
    \left\langle\psi\left|
        \left[H_B^{\mathrm{NC}},H_C\right]
    \right|\psi\right\rangle
    \right|
    =
    \left|
    \left\langle\psi\left|
        \sum_{r=1}^{M/2}A_r^{\mathrm{NC}}
    \right|\psi\right\rangle
    \right|
    =
    2\sqrt{2}\,\frac{M}{2}
    =
    \sqrt{2}\,M.
\end{align}
On the other hand, the triangle inequality gives
\begin{align}
    \left\|
        \left[H_B^{\mathrm{NC}},H_C\right]
    \right\| \leq
    \sum_{r=1}^{M/2}
    \left\|A_r^{\mathrm{NC}}\right\|
    =
    2\sqrt{2}\,\frac{M}{2}
    =
    \sqrt{2}\,M.
\end{align}
Therefore,
\begin{equation}
    \boxed{
    \left\|
        \left[H_B^{\mathrm{NC}},H_C\right]
    \right\|
    =
    \sqrt{2}\,M.
    }
\end{equation}

Finally, for both constructions, $g_B=2,\   k_C^{\mathrm{eff}}=1, $ and $ \  \|H_C\|   =  \sqrt{\frac{M}{2}},$ and hence
\begin{equation}
    2g_Bk_C^{\mathrm{eff}}\|H_C\|
    =
    2\sqrt{2M}.
\end{equation}
The commuting construction therefore exactly saturates the
commuting-battery bound,
\begin{equation}
    \left\|
        \left[H_B^{\mathrm C},H_C\right]
    \right\|
    =
    2g_Bk_C^{\mathrm{eff}}\|H_C\|.
\end{equation}
In contrast, for the non-commuting construction,
\begin{equation}
    \frac{
        \left\|
            \left[H_B^{\mathrm{NC}},H_C\right]
        \right\|
    }{
        2g_Bk_C^{\mathrm{eff}}\|H_C\|
    }
    =
    \frac{\sqrt{M}}{2}.
\end{equation}
Thus, for $M>4$, the non-commuting construction exceeds the commuting-battery bound, with a violation that grows as $\sqrt{M}$.

\section*{Exact transition structure in the battery-energy basis}
\label{sec:SM_AKLH_blocks}

We now determine the projected charging blocks
\begin{equation}
    \Pi_{E'}^B H_C \Pi_E^B
\end{equation}
for both the commuting and noncommuting batteries.  This makes the energy transitions discussed in the main text explicit and provides the exact quantities used in the AKLH transition plots.

\subsection*{Commuting battery}

For the commuting battery
\(
H_B^{\mathrm C}=\sum_{a=1}^{M}P_a^{\mathrm C}
\),
let \(p_a=\pm1\) denote the eigenvalue of \(P_a^{\mathrm C}\).
The global constraint \(\prod_{a=1}^{M}p_a=1\) requires the
numbers \(n_\pm\) of positive and negative eigenvalues to be even.
Consequently,
\begin{equation}
    E=n_+-n_-=2n_+-M,
    \qquad
    E=-M,-M+4,\ldots,M.
\end{equation}

Write \(m:=M/2\), denote the matching edges by
\(e_r=(2r-1,2r)\), and let \(Q_r=\chi_C \gamma_rY_{e_r}\).
The operator \(Q_r\) anticommutes with
\(P_{2r-1}^{\mathrm C}\) and \(P_{2r}^{\mathrm C}\), and commutes
with every other battery term. Define
\begin{align}
    \pi_r^{++}
    &=
    \frac{1}{4}
    (\mathbb I+P_{2r-1}^{\mathrm C})
    (\mathbb I+P_{2r}^{\mathrm C}),\\
    \pi_r^{--}
    &=
    \frac{1}{4}
    (\mathbb I-P_{2r-1}^{\mathrm C})
    (\mathbb I-P_{2r}^{\mathrm C}),\\
    \pi_r^{\mathrm{mix}}
    &=
    \frac{1}{2}
    (\mathbb I-P_{2r-1}^{\mathrm C}P_{2r}^{\mathrm C}).
\end{align}
These projectors satisfy
\begin{equation}
    \mathbb I
    =
    \pi_r^{++}+\pi_r^{--}+\pi_r^{\mathrm{mix}},
\end{equation}
and
\begin{equation}
    Q_r\pi_r^{++}=\pi_r^{--}Q_r,
    \qquad
    Q_r\pi_r^{--}=\pi_r^{++}Q_r,
    \qquad
    [Q_r,\pi_r^{\mathrm{mix}}]=0.
\end{equation}

We first consider the \(\Delta E=-4\) block,
\begin{equation}
    B_{E\rightarrow E-4}
    :=
    \Pi_{E-4}^{\mathrm C}H_C\Pi_E^{\mathrm C}
    =
    \Pi_{E-4}^{\mathrm C}
    \sum_{r=1}^{m}Q_r\pi_r^{++}
    \Pi_E^{\mathrm C}.
\end{equation}
To evaluate its norm, consider the invariant subspace $\mathcal K_0  :=   \left\{  |\psi\rangle:   \pi_r^{\mathrm{mix}}|\psi\rangle=0 \ \forall \ r  \right\}$, in which every matching pair is either \(++\) or \(--\). On this subspace, define
\begin{equation}
    \ell_r:=Q_r\pi_r^{++}
    =\pi_r^{--}Q_r.
\end{equation}
Using \(Q_r^2=\mathbb I\) and
\(\{Q_r,Q_s\}=0\) for \(r\neq s\), one obtains
\begin{equation}
    \{\ell_r,\ell_s\}=0,
    \qquad
    \{\ell_r,\ell_s^\dagger\}
    =
    \delta_{rs}\mathbb I
    \quad\text{on }\mathcal K_0.
\end{equation}
Let $L=\sum_r l_r$. Then $L^2=0$ \& $\{L,L^\dagger\}=m\mathbb{I}$ $\Rightarrow$ $(L^\dagger L)^2=L^\dagger LL^\dagger L=L^\dagger(-L^\dagger L+m\mathbb{I})L=mL^\dagger L$. Therefore $ \|L\|=\sqrt{m}$ or $0$, and since $L$ lowers $++$ pairs at energy $E$ to $--$ at energy $E-4$ on the perfect matching, we get
\begin{equation}
    \|\Pi_{E-4}H_C\Pi_E\|=\sqrt{m}.
\end{equation}
Hence
\begin{equation}
    \boxed{
    \left\|
        \Pi_{E-4}^{\mathrm C}H_C\Pi_E^{\mathrm C}
    \right\|
    =
    \sqrt{\frac{M}{2}}
    },
    \qquad
    E=-M+4,\ldots,M.
\end{equation}
By Hermiticity,
\begin{equation}
    \boxed{
    \left\|
        \Pi_{E+4}^{\mathrm C}H_C\Pi_E^{\mathrm C}
    \right\|
    =
    \sqrt{\frac{M}{2}}
    },
    \qquad
    E=-M,\ldots,M-4.
\end{equation}

Finally, the \(\Delta E=0\) block is
\begin{equation}
    D_E
    :=
    \Pi_E^{\mathrm C}H_C\Pi_E^{\mathrm C}
    =
    \Pi_E^{\mathrm C}
    \sum_{r=1}^{m}Q_r\pi_r^{\mathrm{mix}}
    \Pi_E^{\mathrm C}.
\end{equation}
The operators \(Q_r\pi_r^{\mathrm{mix}}\) mutually anticommute and
satisfy
\(
(Q_r\pi_r^{\mathrm{mix}})^2=\pi_r^{\mathrm{mix}}
\).
Therefore,
\begin{equation}
    D_E^2
    =
    \Pi_E^{\mathrm C}
    \left(\sum_{r=1}^{m}\pi_r^{\mathrm{mix}}\right)
    \Pi_E^{\mathrm C}.
\end{equation}
At energy \(E\), the maximum possible number of mixed matching
pairs is
\begin{equation}
    \min(n_+,n_-)
    =
    \frac{M-|E|}{2}.
\end{equation}
It follows that
\begin{equation}
    \boxed{
    \left\|
        \Pi_E^{\mathrm C}H_C\Pi_E^{\mathrm C}
    \right\|
    =
    \sqrt{\frac{M-|E|}{2}}
    }.
\end{equation}
All remaining projected blocks vanish.

\subsection*{Noncommuting battery}
For the non-commuting battery, $(H_B^{\mathrm{NC}})^2=M\mathbb{I}$ and $\operatorname{Tr}(H_B^{\mathrm{NC}})=0$. Thus, $H_B^{\mathrm{NC}}=\sqrt{M}(\Pi_+-\Pi_-)$, where $\Pi_+$ and $\Pi_-$ project onto the high- and low-energy sectors, respectively, and have equal ranks.

Writing
\[
H_C=
\begin{pmatrix}
A & B\\
B^\dagger & D
\end{pmatrix},
\]
where $A=\Pi_+H_C\Pi_+$, $B=\Pi_+H_C\Pi_-$, and $D=\Pi_-H_C\Pi_-$, yields the commutator
\[
[H_B^{\mathrm{NC}},H_C]
=
2\sqrt{M}
\begin{pmatrix}
0 & B\\
-B^\dagger & 0
\end{pmatrix}.
\]
Thus, $\|[H_B^{\mathrm{NC}},H_C]\|=2\sqrt{M}\|B\|$, and hence $\|B\|=\|[H_B^{\mathrm{NC}},H_C]\|/(2\sqrt{M})=\sqrt{M/2}$.

Since $\|\Pi_+H_C\Pi_-\|=\|B\|=\sqrt{M/2}$, there are matrix elements of $H_C$ connecting the battery energies $E_\pm=\pm\sqrt{M}$, which are separated by $|E_+-E_-|=2\sqrt{M}$.

Similarly, the anticommutator yields
\[
\{H_B^{\mathrm{NC}},H_C\}
=
2\sqrt{M}
\begin{pmatrix}
A & 0\\
0 & -D
\end{pmatrix}.
\]
For $M=0\pmod 4$, we can also obtain the diagonal block norms directly. As shown earlier, the commutator can be written as
\[
[H_B^{\mathrm{NC}},H_C]
=
-2i\sqrt{2}\sum_{r=1}^{M/2}S_r,
\]
where the $S_r$ commute with one another, satisfy $(S_r)^2=\mathbb{I}$, and have eigenvalues $\pm1$. Since $M/2$ is even, we may choose a common eigensector with half of the $S_r$ equal to $+1$ and half equal to $-1$. Hence the commutator vanishes on this nonzero subspace. Since $(H_B^{\mathrm{NC}})^2=M\mathbb{I}$ and $H_C^2=(M/2)\mathbb{I}$, $H_B^{\mathrm{NC}}$ and $H_C$ can be simultaneously diagonalized within this subspace, with eigenvalues $\pm\sqrt{M}$ and $\pm\sqrt{M/2}$, respectively. Finally, $\Omega=\prod_{a=1}^{M}P_a^{\mathrm{NC}}$ anticommutes with $H_B^{\mathrm{NC}}$ and commutes with $H_C$, so a simultaneous eigenstate in one battery-energy sector is mapped to one in the other sector with the same $H_C$ eigenvalue. Therefore,
\[
\|\Pi_+H_C\Pi_+\|
=
\|\Pi_-H_C\Pi_-\|
=
\sqrt{\frac{M}{2}}.
\]
Whereas the off-diagonal result $\|\Pi_+H_C\Pi_-\|=\sqrt{M/2}$ holds for every $M$, giving a growing energy-transition width with system size.

\subsection*{Comparison of transitions}

The contrast between the two battery algebras is now explicit.  For the commuting battery case, $ \Pi_{E'}^{\rm C}H_C\Pi_E^{\rm C}=0 \  \text{for}  \  |E'-E|>4,$ in agreement with
\begin{equation}
    2\mathcal G_{B|C}
    =
    2g_Bk_C^{\rm eff}
    =
    4.
\end{equation}
For the noncommuting battery, however, $\left\|\Pi_+H_C\Pi_- \right\| =\sqrt{\frac M2}$ at an energy separation $|\Delta E|  =  2\sqrt M.$ The ratio between this accessible energy separation and the commuting transition width is therefore $\frac{2\sqrt M}{4}  = \frac{\sqrt M}{2},$ which diverges with system size.  The AKLH transition plots in the main text are obtained directly from the exact expressions above.

\section*{Exact joint ground-state sector charging dynamics}
\label{sec:SM_dynamics}

We show here that a suitable joint state supported entirely in the battery ground-energy sector exists whose charging power achieves the commutator norm dynamically. We have $(H_B^{\mathrm{NC}})^2=M\mathbb{I}$ and $H_C^2=m\mathbb{I}$, where $m=M/2$. Writing $H_B^{\mathrm{NC}}=\sqrt{M}(\Pi_+-\Pi_-)$, where $\Pi_+$ is the projector onto the excited-state manifold and $\Pi_-$ the projector onto the ground-state manifold, the time-evolution operator is
\begin{equation}
    U_C(t)=e^{-iH_Ct}
    =\cos(\sqrt{m}t)\mathbb{I}
    -i\frac{\sin(\sqrt{m}t)}{\sqrt{m}}H_C.
\end{equation}

From earlier, we know that the charging block satisfies $\|\Pi_+H_C\Pi_-\|=\sqrt{m}=\|H_C\|$. By definition of the operator norm, there exists a normalized joint vector $\Pi_-|g_*\rangle=|g_*\rangle$ such that $\|\Pi_+H_C\Pi_-|g_*\rangle\|=\sqrt{m}$. Also, since $H_C^2=m\mathbb{I}$, $\langle g_*|H_C^2|g_*\rangle=m$, and hence $\|H_C|g_*\rangle\|^2=m$. Resolving this into the two battery-energy sectors $\Pi_+$ and $\Pi_-$ gives $\|H_C|g_*\rangle\|^2=\|\Pi_+H_C|g_*\rangle\|^2+\|\Pi_-H_C|g_*\rangle\|^2=m$. Therefore $\|\Pi_-H_C|g_*\rangle\|^2=0$. In other words, $H_C|g_*\rangle$ is a joint state supported entirely in the battery excited-energy sector. Define $|e_*\rangle=H_C|g_*\rangle/\sqrt{m}$. We then have $H_B^{\mathrm{NC}}|e_*\rangle=\sqrt{M}|e_*\rangle$.

Hence the two-dimensional subspace $\mathcal{H}_*=\operatorname{span}\{|g_*\rangle,|e_*\rangle\}$ is invariant under both $H_B^{\mathrm{NC}}$ and $H_C$, with
\begin{equation}
    \left.H_B^{\mathrm{NC}}\right|_{\mathcal{H}_*}
    =
    \sqrt{M}
    \begin{pmatrix}
        -1 & 0\\
        0 & 1
    \end{pmatrix},
    \qquad
    \left.H_C\right|_{\mathcal{H}_*}
    =
    \sqrt{m}
    \begin{pmatrix}
        0 & 1\\
        1 & 0
    \end{pmatrix}.
\end{equation}

The time evolution of $|g_*\rangle$ is
\begin{equation}
    U_C(t)|g_*\rangle
    =
    \cos(\sqrt{m}t)|g_*\rangle
    -i\sin(\sqrt{m}t)|e_*\rangle.
\end{equation}
With stored battery energy $E_B(t)=\langle H_B^{\mathrm{NC}}\rangle_t$, we obtain $E_B(t)=-\sqrt{M}\cos(2\sqrt{m}t)$, giving
\begin{equation}
    P(t)
    =
    2\sqrt{Mm}\sin(2\sqrt{m}t)
    =
    \sqrt{2}\,M\sin(2\sqrt{m}t),
\end{equation}
and hence
\begin{equation}
    \max_t |P(t)|
    =
    \sqrt{2}\,M
    =
    \|[H_B^{\mathrm{NC}},H_C]\|.
\end{equation}

\subsection*{State selectivity within the joint ground-state sector manifold}

The previous result establishes the existence of a suitable joint ground state $|g_\star\rangle$, but the ground state manifold of $H_B^{\rm NC}$ is exponentially degenerate with dimension  $\dim{\cal G}_B   =  2^{N-1}$, and the joint state also includes the auxiliary Majorana Hilbert space. A generic ground-state vector need not saturate the norm of the commutator.

For an arbitrary normalized joint state $|g\rangle$, define
\begin{equation}
    w_g
    :=
    \frac{1}{m}
    \left\|
    \Pi_+H_C|g\rangle
    \right\|^2,
    \qquad
    0\leq w_g\leq1.
    \label{eq:wg_def}
\end{equation}
Since  $\|H_C|g\rangle\|^2=m,$ the complementary weight remaining in the battery ground-states energy sector is
\begin{equation}
    \frac{1}{m}
    \left\|
    \Pi_-H_C|g\rangle
    \right\|^2
    =
    1-w_g.
\end{equation}
Substitution into the exact evolution gives
\begin{equation}
    E_B^{(g)}(t)
    =
    -\sqrt M
    +
    2\sqrt M\,
    w_g\sin^2(\sqrt m\,t),
    \label{eq:general_ground_energy}
\end{equation}
and hence $ P_g(t)  = 2\sqrt{mM}\,w_g\sin(2\sqrt m\,t),$ yielding $ P_{\max}^{(g)} = \sqrt2\,M\,w_g.$ The optimal state has $w_{g_\star}=1, $ and saturates the commutator norm, whereas other joint vectors in the degenerate ground state manifold can exhibit different power scalings.

\end{document}